\documentclass[12pt]{spieman}  
\usepackage{amsmath,amsfonts,amssymb}
\usepackage{graphicx}
\usepackage{setspace}
\usepackage{tocloft}

\usepackage{ulem}

\usepackage{color,soul}
\usepackage{float}
\usepackage{booktabs}
\usepackage{tabularx} 
\usepackage[normal]{threeparttable}
\usepackage{xfrac} 
\usepackage{colortbl} 
\definecolor{Gray}{gray}{0.85} 
\usepackage{multirow}
\usepackage[makeroom]{cancel} 
\usepackage{footmisc}

\title{Proton irradiation results for long-wave HgCdTe infrared detector arrays for NEOCam}

\author[a,*]{M. Dorn}
\author[a]{J. L. Pipher}
\author[a]{C. McMurtry}
\author[b]{S. Hartman}
\author[c]{A. Mainzer}
\author[d]{M. McKelvey}
\author[d]{R. McMurray}
\author[a]{D. Chevara}
\author[a]{J. Rosser}
\affil[a]{University of Rochester, Department of Physics and Astronomy, Rochester, NY, 14627}
\affil[b]{University of California Davis, Crocker Nuclear Laboratory, Davis, CA, 95616}
\affil[c]{Jet Propulsion Laboratory, California Institute of Technology, Pasadena, CA, 91125}
\affil[d]{Ames Research Center, Moffett Field, CA, 94035}

\cftpagenumbersoff{figure}
\cftpagenumbersoff{table} 
\begin{document} 
\maketitle

\begin{abstract}
HgCdTe detector arrays with a cutoff wavelength of $\sim$10 $\mu$m intended for the NEOCam space mission were subjected to proton beam irradiation at the University of California Davis Crocker Nuclear Laboratory. Three arrays were tested - one with 800 $\mu$m substrate intact, one with 30 $\mu$m substrate, and one completely substrate-removed. The CdZnTe substrate, on which the HgCdTe detector is grown, has been shown to produce luminescence in shorter wave HgCdTe arrays that causes elevated signal in non-hit pixels when subjected to proton irradiation. This testing was conducted to ascertain whether or not full substrate removal is necessary. At the dark level of the dewar, we detect no luminescence in non-hit pixels during proton testing for both the substrate-removed detector array and the array with 30 $\mu$m substrate. The detector array with full 800 $\mu$m substrate exhibited substantial photocurrent for a flux of 103 protons/cm$^2$-s at a beam energy of 18.1 MeV ($\sim$ 750 e$^-$/s) and 34.4 MeV ($\sim$ 65 e$^-$/s). For the integrated space-like ambient proton flux level measured by the Spitzer Space Telescope, the luminescence would be well below the NEOCam dark current requirement of \textless200 e$^-$/s, but the pattern of luminescence could be problematic, possibly complicating calibration.
\end{abstract}

\keywords{infrared detectors, infrared, detector array}

{\noindent \footnotesize\textbf{*}E-mail:  \linkable{meghan@pas.rochester.edu} }

\begin{spacing}{2}   

\section{Introduction}
\label{sect:intro} 

We have been developing sensitive long-wave infrared (LWIR) 10 $\mu$m cutoff Hg$_{1-x}$Cd$_{x}$Te detector arrays for use in the proposed NEOCam Discovery mission.\cite{Mainzer15}\footnote[2]{Hg$_{1-x}$Cd$_{x}$Te (mercury cadmium telluride) is a II-VI ternary compound, whose molar cadmium fraction, $x$, can be varied to tune to the desired cutoff wavelength.} NEOCam (Near Earth Object Camera) is a survey mission designed to find, track, and characterize asteroids and comets in our solar system, including most of those greater than 140 m in size that travel close to Earth, the Near Earth Objects (NEOs). There is particular emphasis on finding those NEOs with the potential of impacting the Earth. NEOCam will operate at two wavelength ranges - 4 to 5 $\mu$m and 6 to 10 $\mu$m. The shorter wavelength HgCdTe detector arrays have already been designed for the James Webb Space Telescope (JWST)\cite{Rauscher2014}, and our group at The University of Rochester (UR), in collaboration with NASA JPL, has been working with Teledyne Imaging Sensors (TIS) to produce the detector arrays that will cover the longer wavelength range. Laboratory testing of several arrays grown on an 800 $\mu$m CdZnTe substrate, and hybridized to an H1RG multiplexer, demonstrated that the resulting detector arrays met all NEOCam requirements (see Table~\ref{table:neocamrequirements}) for dark current, quantum efficiency, well depth and noise.\cite{McMurtry2013} Short-wave infrared (SWIR) and mid-wave infrared (MWIR) HgCdTe detector arrays utilizing the same multiplexer, or the same family of multiplexers in a larger format, have been or will be employed in other space missions, including the Orbiting Carbon Observatory 2, the Wide-field Infrared Survey Explorer, the Hubble Space TelescopeÕs Wide Field Camera 3, Euclid, and JWST.\cite{Oday2011,Mainzer2008,Baggett2008, Rauscher2014, Laureijs2012} Detector arrays flown in space must be robust against cosmic ray hits; therefore, we subjected the arrays to 12 MeV-63 MeV protons to determine the magnitude of the responses.

\begin{table}[H]
\centering
\caption{Minimum NEOCam Requirements and Detector Array Characteristics}
\begin{threeparttable}
\begin{tabular}{lccccc} \toprule
   & \multicolumn{1}{c}{\textbf{NEOCam}} & \multirow{ 2}{*}{\textbf{Goal}} & \multicolumn{1}{c}{\textbf{H1RG-}} & \multicolumn{1}{c}{\textbf{H1RG-}} & \multicolumn{1}{c}{\textbf{H1RG-}} \\ 
   & \multicolumn{1}{c}{\textbf{Requirement }} &  & \multicolumn{1}{c}{\textbf{17346}} & \multicolumn{1}{c}{\textbf{17354}} & \multicolumn{1}{c}{\textbf{16886}} \\ 
   \midrule
 \rowcolor{Gray}
Array Format & 1024x1024 & 2048x2048  & 1024x1024 & 1024x1024 & 1024x1024 \\ 
Cutoff Wavelength & \multirow{ 2}{*}{10} & \multirow{ 2}{*}{10.3} & \multirow{ 2}{*}{9.9}  & \multirow{ 2}{*}{9.9} & \multirow{ 2}{*}{10.2}  \\ 
($\mu$m)  &   &   &   &   &  \\
\rowcolor{Gray}
Responsive Quantum & 55 & 55 & 58 & 56 & 58 \\
\rowcolor{Gray}
Efficiency (RQE) (\%) & (Non-ARC) & (Non-ARC) & (Non-ARC) & (Non-ARC) & (Non-ARC) \\
CDS read noise (e$^-$) & 36 & 30 & 22 & 22 & 22 \\
\rowcolor{Gray}
Dark Current (e$^-$/s) & \textless 200 & \textless 1 & \textless 1* & \textless 1* & \textless 1* \\
Well Depth (e$^-$) & \textgreater 46,000 & \textgreater 65,000 & \textgreater 72,400 & \textgreater 64,600 & \textgreater 57,300 \\
\rowcolor{Gray}
Operability (\%) & 90 & 90 & 93.7 & 93.5 & 96.7 \\
\bottomrule
\end{tabular} 
\begin{tablenotes}
      \small
      \item ARC = Anti Reflection Coating
      \item *The dark current value quoted is the median and histograms are shown in Appendix A.
\end{tablenotes}
\end{threeparttable}
\label{table:neocamrequirements}   
\end{table}

\section{Motivation}
\label{sect:motivation} 

Waczynski et al. (2005) showed that SWIR HgCdTe arrays intended for the HST Wide Field Camera 3 instrument exhibited an elevated signal level in ÒbackgroundÓ (non-hit) pixels during 15.7-63 MeV proton irradiation. They found evidence that the spatial distribution of the elevated signal level across the array is correlated with the responsivity to flood illumination with 800 nm light, and the elevated signal level is proportional to some fraction of the proton energy deposited in the CdZnTe substrate. Energy deposited in the 800 $\mu$m thick substrate can create electron hole pairs, some of which may radiatively recombine and emit 775 nm radiation, corresponding to the bandgap of the CdZnTe.\cite{quijada2007}

The observed luminescence has been eliminated on SWIR arrays by removing the CdZnTe substrate.\cite{Piquette2008,Waczynski2005,Smith2006} MWIR arrays developed for the James Webb Space Telescope (JWST) were also substrate-removed by TIS, utilizing the same process used for SWIR arrays.\cite{Garnett2004} In order to remove the substrate, the volume between the silicon multiplexer and the detector must be epoxy backfilled, because the light-sensitive HgCdTe layer is thin ($\sim$ 10 $\mu$m). McMurtry et al. (2013) investigated the effect of epoxy backfill on the substrate-intact LWIR devices and showed that it did not adversely affect the dark current, well depth, quantum efficiency, or noise although an increase in interpixel capacitance was noted. 

The first epoxy-backfilled and substrate-removed array produced for this project was also the first substrate-removed LWIR array produced by TIS. This array exhibited excellent dark current. However, 99.7\% of pixels had well depth below the NEOCam requirement (second column, Table~\ref{table:neocamrequirements}), and the quantum efficiency was extremely low. Following the production of this first substrate-removed array, we received another substrate-removed array, H1RG-17346. A quarter of the array met NEOCam requirements. Subsequently, TIS produced two more substrate-removed arrays that also did not meet NEOCam requirements.

The poor overall performance of the substrate-removed devices tested, and the low yield of the LWIR substrate removal process, motivated the question as to whether complete substrate removal was necessary. Consequently, we pursued the fabrication of a device with most, but not all, of its substrate removed. H1RG-17354 was delivered with approximately 30 $\mu$m substrate remaining. H1RG-17354 exhibited excellent dark current, quantum efficiency and noise characteristics and met all NEOCam requirements. However, we were concerned that the remaining 30 $\mu$m substrate would produce some degree of `background' luminescence under cosmic ray irradiation. To mitigate this concern, we subjected the one successfully substrate-removed device (H1RG-17346), the partially substrate-removed device (H1RG-17354), and a fully substrate intact device (H1RG-16886) to 12 - 63 MeV proton irradiation at the University of California Davis Crocker Nuclear Laboratory (CNL) cyclotron. Previous tests in our lab at UR found no evidence for any substrate luminescence from cosmic ray hits, mostly 4 GeV muons, for any of the 10 $\mu$m cutoff arrays satisfying the NEOcam requirements (Table~\ref{table:neocamrequirements}), including the arrays discussed in this paper.\cite{Girard2014} 


\section{Proton Stopping Power}
\label{theory}

\noindent
High-energy protons lose energy as they travel through both the CdZnTe substrate and the HgCdTe detector layer. The energy loss can generate charge carriers via ionization. The energy loss for charged heavy particles traveling through a material is calculated to first order by the Bethe-Bloch formula\cite{Nakamura2010,Bethe1930}

\begin{equation}\label{eq:energytransfer}
- \langle dE/dx \rangle = Kz^2\frac{Z}{A}\frac{1}{\beta^2} \bigg[ \frac{1}{2}ln\frac{2m_ec^2\beta^2\gamma^2T_{max}}{I^2}-\beta^2\bigg]\rho
\end{equation}

\noindent
where $m_e$ is the mass of the electron, $c$ is the speed of light, $z$ is the charge on the particle (proton charge, $z=+1$), $Z$ is the effective atomic number of the material, $A$ is the atomic mass, $T_{max}$ is the maximum energy transfer in a single collision, $I$ is the mean excitation energy, and $\rho$ is the density of the material. The constant $K$ is given by

\begin{equation}\label{eq:Kconstant}
K = 4\pi N_A r_e^2m_ec^2
\end{equation}

\noindent
where $N_A$ is Avogadro's number, and $r_e$ is the radius of an electron. The Bethe-Bloch  formula describes the mean energy loss/unit length  for 0.1 $\lesssim \beta\gamma \lesssim$ 1000, where $\beta$ is the ratio of velocity to speed of light in vacuum, and $\gamma$ is the Lorentz factor of the particle. The equation can also include a density correction, not included here, as it is not relevant. 

Generally, relatively high-energy particles lose less energy to the medium and have a greater range, while relatively low-energy particles may lose enough energy to stop within the medium and thus typically cause more damage or upsets to electronic components. As high-energy protons traverse the detector, the liberated holes are collected by the diode as a current component. The effective ionization potential for long-wavelength cutoff HgCdTe is $\sim$ 1.04 eV/e$^-$, corresponding to an efficiency of converting energy into electron-hole pairs of about 10\% (the band gap energy is about 0.1 eV).\cite{Marshall2003} The total charge generated by high-energy protons is approximately\cite{Pickel2004}

\begin{equation}\label{eq:chargegen}
\mbox{charge} \approx \mbox{pathlength * ($dE/dx$) / ionization energy per e$^-$}
\end{equation}

\noindent
For this calculation, we assume normal incidence. Luminescence from the CdZnTe substrate is another consequence of ionization by lower energy protons  (of the energy range tested). The most damaging energy range of protons is $\sim$ 0.1-20 MeV, since \textless 20 MeV protons can stop in the substrate or detector material. The Bragg peak resides within the 30 $\mu$m substrate for proton energies \textless 4.8 MeV and within the detector material for energies between 4.8 - 5.1 MeV. For arrays with full 800 $\mu$m substrate, the Bragg peak resides just within the substrate for 15.7 MeV protons.\cite{Waczynski2005} Although the cosmic ray spectrum is composed primarily of protons, alpha particles in the range of $\sim$ 0.1-20 MeV have approximately the same flux as protons in that range, and can be more damaging, because they are stopped much more efficiently than protons.\cite{Logachev2002,Simpson1983} The energy transfer, $dE/dx$, (Eqn.~\ref{eq:energytransfer}), is directly proportional to $z^2$, where $z$ is the particle charge: therefore an alpha particle of the same energy as a proton would be stopped in approximately one quarter of the distance.

\begin{figure} [H]
\begin{center}
\begin{tabular}{c}
\includegraphics[height=5.5cm]{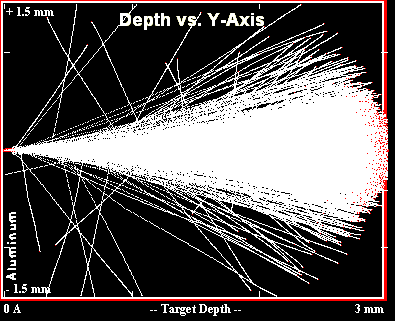}
\hfill
\includegraphics[height=5.5cm]{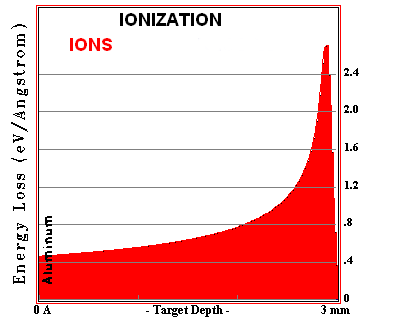}
\hspace{1cm}
\end{tabular}
\end{center}
\caption 
{ \label{fig:srimaluminum}
Left: Simulated proton tracks through 3 mm of aluminum. Right: Energy loss in aluminum as a function of target thickness for 24 MeV protons. Simulation performed using SRIM.} 
\end{figure}

For a typical spacecraft, some shielding will be present.  This shielding tends to stop the relatively low-energy component of the expected cosmic ray spectrum, which consists of galactic cosmic rays and solar quiet-time particles, while simultaneously degrading the energy of the remaining higher energy cosmic rays.\cite{Logachev2002}  For 3 mm of Al shielding in all directions, all protons with energies below $\sim$ 24 MeV will be stopped (See Fig.~\ref{fig:srimaluminum}, calculated using SRIM, the Stopping and Range of Ions in Matter, a selection of software packages which can be downloaded from srim.org that calculate many features of the transport of ions in matter.  A textbook by the same name describes the methodology and application.)\cite{srim, srimbook} For the same amount of shielding, alpha particles with energy below 95.75 MeV will be stopped. The peak of the alpha particle component of the cosmic ray spectrum occurs at the same energy as that of the proton spectrum, but the flux of alpha particles not stopped by the shielding is 10 times lower than that of the protons.

\noindent
The cosmic protons with initial energies between $\sim$ 24 - 32 MeV will degrade to 0.1 - 18 MeV after passage through 3 mm of aluminum (from a TRIM calculation: TRIM, TRansport of Ions In Matter, is a program in the SRIM package that calculates the energy loss of ions moving through matter using a quantum mechanical treatment of atom-atom collisions). These lower-energy, damaging cosmic rays are experienced by a spacecraft during the solar storms emanating from the Sun after strong flares lead to coronal mass ejection events, but detectors are exposed to an approximately steady differential flux at energies \textless 18 MeV (14\% of the peak galactic cosmic ray differential flux at 300 MeV). Higher energy (\textgreater 28 MeV) protons and cosmic rays will tend to pass directly through the shielding as well as the detector and substrate material and will deposit less energy than lower energy protons in the detector material. The IRAC InSb and Si:As detector arrays on the Spitzer Space Telescope, with similar aluminum shielding, experienced transient rates of pixels hit ranging from 3-10 s$^{-1}$ over the course of the cryogenic mission, excepting solar flares, which led to much higher rates.\cite{hora2006}

\section{Experimental Tests and Methods}
\label{sect:testsandmethods}  

The primary objectives of the proton radiation tests are to
\begin{enumerate}
\item	Characterize the spatial extent and energy deposited by isolated proton hits;
\item	Examine any residual effects and recovery time from proton hits;
\item	Investigate the dark current in non-hit pixels during proton irradiation;
\item	Assess array performance following a cumulative lifetime dose
\end{enumerate}

\noindent
The three detector arrays discussed in this paper are 1024x1024x18 $\mu$m pixel pitch LWIR HgCdTe infrared arrays on H1RG read-outs and were tested at CNL. The pixel design is very similar to that of the 2.5 and 5 $\mu$m HgCdTe TIS arrays, except that it is a 10 $\mu$m cutoff. Over several trips, H1RG-17346 and H1RG-17354 were irradiated with 63 MeV, 32 MeV, and 12 MeV protons at various fluxes. H1RG-16886 was irradiated with 34.4 MeV and 18.1 MeV protons. Intermediate metal foils of various thicknesses were used to attenuate the proton beam to the desired energy. The beam energies quoted were measured at the exit of the cyclotron. Table~\ref{table:dataruns} summarizes the test dates, the array tested, and the nature of the tests, as well as the beam energies utilized. Prior to the CNL proton tests, similar data were obtained in the Ames lab and at CNL without the proton beam irradiating the array under test, in the same test dewar.

We made use of the NASA Ames array controller and acquisition system, as well as their test dewar, since these have been extensively used for other proton irradiation tests for space experiments, including JWST, Spitzer, WISE, and HST among others.  The dewar and experimental setup at CNL has been described in detail by others.\cite{Waczynski2005} A 5 mil Kapton window at a side entrance of the dewar passed the beam through three radiation shield windows masked by 1 mil aluminum foil.  An external aluminum aperture defined the proton beam size to $\sim$ 37mm x 37mm, exceeding the size of the detector array surface being irradiated (\textless 25mm x 25mm).  Normal incidence was used for all tests, with the dewar window situated close to the cyclotron exit beam. The metal windows and masks in the beam path reduced the beam energies incident on the detector array during the first two runs by 0.4 MeV for 63 MeV initial beam energy, by 0.6 MeV for 32 MeV initial beam energy and by 1.3 MeV for 12 MeV initial beam energy (from SRIM simulations). The 18.1 MeV and 34.4 MeV beam energies used in the tests on H1RG-16886 were degraded by 1 MeV and 0.6 MeV respectively. Also, the tested beam energies had a spread in energy due to the accelerator tuning process, and passage through both our windows and masks; e.g. the 12 MeV beam energy was both degraded and spread in energy such that the incident energies on our detector array under test were actually in the range of 10.57-10.93 MeV. 

All data were obtained in sample-up-the-ramp (SUTR) mode at 2 reverse biases (150 mV and 250 mV).  Preliminary dark tests at Ames prior to the trip to CNL, as well as data obtained during proton testing at CNL were obtained primarily in two modes: 4 repetitions with 16 samples per image and 2 repetitions with 64 samples per image. The initial proton tests were obtained at low flux: the minimum calibrated level at CNL is $\sim$ 10$^5$ protons/cm$^2$-s. CNL lab scientists can reduce the level reliably by a factor of 10 to $\sim$ 10$^4$ protons/cm$^2$-s, although dosimetry is unavailable.  Calibration at lower flux levels than 10$^5$ protons/cm$^2$-s must be done through analysis of hits on the array itself by reducing the flux until a small number of proton hits are shown in an image.  A focal plane temperature of 35 K was maintained throughout the experiments. We have shown that the detectors operate well between 30 K - 42 K, and \textless 40 K is the nominal temperature for the NEOCam 10 $\mu$m focal plane (See Fig. 5 in Ref~\citenum{McMurtry2013}). Pixel voltages read out are converted to electrons by using our lab conversion factor, given by the typical nodal capacitance of 42 fF, normalized by the relative gains of the Ames and UR readout electronics. \cite{McMurtry2013}

\begin{table}[H]
\centering
\caption{Summary of data obtained over several radiation runs.}
\begin{tabularx}{\linewidth}{llcX} \toprule
    \multicolumn{1}{c}{\textbf{Date}} & \multicolumn{1}{c}{\textbf{Detector Array}} & \multicolumn{1}{c}{\textbf{Substrate Thickness}}  & \multicolumn{1}{c}{\textbf{Measurements}}                                                   \tabularnewline \midrule
    Sep 26, 2013  & H1RG-17346 & 0 $\mu$m & 7.5 krad(Si) cumulative dose at 63 MeV  \\ \midrule
    Aug 13, 2014  & H1RG-17354 & 30 $\mu$m & Single event data at 63 MeV, 1 krad(Si) and 5 krad(Si) cumulative dose at 63 MeV \\ \midrule
    Sep 4, 2014 & H1RG-17346 & 0 $\mu$m & Single event data for energies 12, 32, and 63 MeV, 5 krad(Si) cumulative dose at 32 MeV \\ \midrule
    Oct 2, 2014 & H1RG-17354 & 30 $\mu$m & Single event data at 12 and 32 MeV \\ \midrule
    Sep 22, 2015 & H1RG-16886 & 800 $\mu$m & Single event data at 18.1 and 34.4 MeV \\ \bottomrule
\end{tabularx} 
\label{table:dataruns}   
\end{table}

\section{Data Analysis}
\label{sect:dataanalysis}  

Dark current and well depth maps for all arrays have previously been produced from measurements made in the lab at the University of Rochester. Three sigma (3$\sigma$) clipping of the bias- and temperature-specific dark current and well depth maps was used to mask out pixels exhibiting either high dark current or low well depth, or both.

For these SUTR data, differences of consecutive frames were used to find pixels whose SUTR data exhibit a \textit{sudden jump} in signal greater than 3x the standard deviation of the average signal from each pixel. These pixels are selected as possible proton hits.

For pixels flagged as a proton hit, a 5x5 box around the flagged pixel was masked off in order to examine the non-hit pixels surrounding the proton strike. We took SUTR data with 5.278 s between samples several times for each flux level and reverse bias level, while resetting the array in between SUTR data sets. Subsequently, the mean and standard deviation of the dark current, the slopes for a steady sample-up-the-ramp, were calculated for non-hit pixels common to all data. The mean `dark current' before irradiation, 0.3 e$^-$/s/pixel, is significantly above the upper limit to dark current, 0.04 e$^-$/s/pixel obtained in the University of Rochester lab.\cite{Girard2014} We believe there is a slight light leak or glow in the Ames dewar; however, since the NEOCam dark current requirement is \textless 200 e$^-$/s, this is an acceptable value.

To investigate the number of pixels affected by a proton hit, pixels previously flagged as a potential hit and a 5x5 region around that pixel were considered. For each transient event, we only considered pixels with a jump in signal 5$\sigma$ above the median background to be part of the proton hit. After the jump, the pixel continues integrating up the ramp, with the same slope as before the jump.

\section{Results and Discussion}
\label{sect:results}  

\subsection{Dark Current Before, During, and After Testing}

The procedure outlined in Section~\ref{sect:dataanalysis} was used to reduce data obtained while the proton beam was tuned to a low flux level, before irradiation, and after. We provide those data for all three arrays below.

\subsubsection{H1RG-17354}
\begin{figure}[H]
\begin{center}
\begin{tabular}{c}
\includegraphics[height=5.5cm]{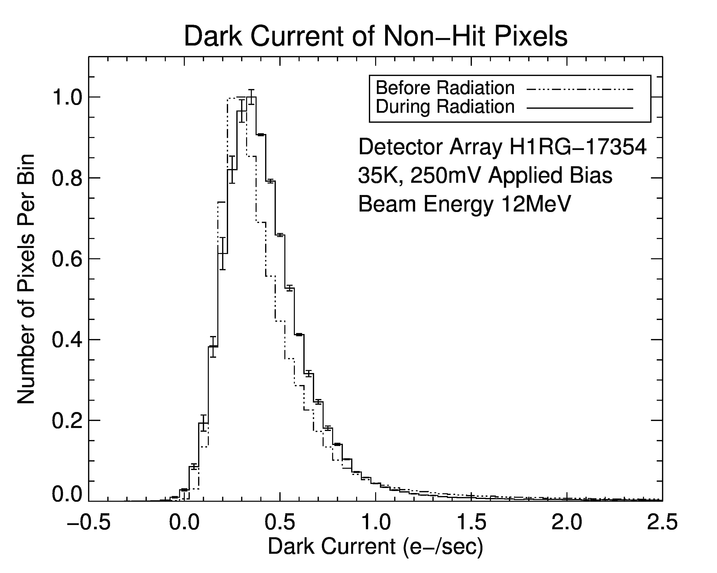}
\hfill
\includegraphics[height=5.5cm]{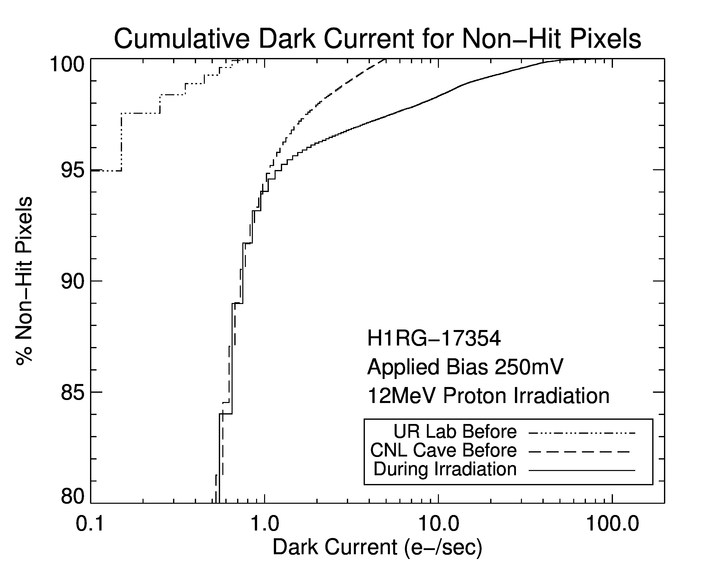}
\hspace{1cm}
\end{tabular}
\end{center}
\caption 
{ \label{fig:darkcurrent354}
Left: Dark current for non-hit pixels ($\sim$130,000 pixels), normalized to peak value, during low dose 12 MeV proton irradiation for H1RG-17354, an array with a 30 $\mu$m substrate. The dark current data obtained at CNL with the beam stop in place before radiation testing are overplotted on the radiation data. Right: Cumulative dark current of non-hit pixels during 12 MeV proton irradiation. Cumulative dark current measured in the lab at University of Rochester, and cumulative dark current measured in the laboratory space outside the beam chamber immediately before irradiation, are overplotted. The applied reverse bias for these data is 250mV.} 
\end{figure}

The first detector array we tested was H1RG-17354, an array with 30 $\mu$m substrate. Figure~\ref{fig:darkcurrent354} shows the dark current and the cumulative dark current immediately before and during irradiation by 12 MeV protons. 

\noindent
Low fluence data for non-hit pixels do not show an appreciable increase in dark current during irradiation compared with dark current data obtained in the laboratory space outside of the beam chamber before irradiation, within the uncertainty in the measurement. Table~\ref{table:darkcurrent354} summarizes the modal dark current for all beam energies at two different applied biases. The mode was computed using a function in IDL that determines the value of an array where the maximum number of elements of the array are located given a bin size. The uncertainty in the measurements is included: it is important to note that within the uncertainties in the dark measurement before and during irradiation, there is consistently no increase in modal dark current for all beam energies (the median yields the same result).

\begin{table}[H]
\centering
\caption{Modal dark current in the laboratory space outside the beam chamber before irradiation and during irradiation for various beam energies for H1RG-17354, an array with a 30 $\mu$m substrate. *No baseline data are available for a 63 MeV beam at applied bias of 150 mV. **Baseline data for a 63 MeV beam were obtained in the lab at NASA Ames in the same dewar.}
\begin{threeparttable}
\begin{tabular}{lll} \toprule
    \multicolumn{1}{c}{\textbf{Beam Energy }} & \multicolumn{2}{c}{\textbf{Modal Dark Current (e$^-$/s/pixel) }}  \\ \cmidrule(r){2-3}
    \multicolumn{1}{c}{\textbf{(MeV)}}              & \multicolumn{1}{c}{\textbf{150 mV Reverse Bias}} & \multicolumn{1}{c}{\textbf{250 mV Reverse Bias}} \\  \midrule
    12$^1$ \textit{(and before irradiation)} & 0.40 $\pm$ 0.12 \textit{(0.23)} & 0.36 $\pm$ 0.09 \textit{(0.32)} \\
    32$^1$ \textit{(and before irradiation)} & 0.41 $\pm$ 0.12 \textit{(0.23)} & 0.46 $\pm$ 0.09 \textit{(0.32)} \\
    63$^2$ \textit{(and before irradiation)} & * & 0.08 $\pm$ 0.04 \textit{(0.04**)} \\ \bottomrule
\end{tabular} 
\begin{tablenotes}
      \small
      \item $^1$Data run 2 Oct 2014
      \item $^2$Data run 13 Aug 2014
\end{tablenotes}
\end{threeparttable}
\label{table:darkcurrent354}   
\end{table}

\noindent
The cumulative dark current for 250mV applied bias is plotted on the right in Fig.~\ref{fig:darkcurrent354} in order to illustrate the dark current levels of non-hit pixels during low dose irradiation. \textbf{All non-hit pixels meet the NEOCam dark current requirement of \textless 200 e$^-$/s/pixel during irradiation.} On the left in Fig.~\ref{fig:darkcurrent354}, a small dark current range of $\sim$ -0.5-2.5 e$^-$/s is shown, with a linear scale on the y-axis. A few pixels exhibit higher current up to $\sim$ 150 e$^-$/s and are not shown on the plot. 
 
From Equation~\ref{eq:energytransfer}, the energy loss in the 30 $\mu$m thick substrate for 12 MeV protons is 394 keV, for 32 MeV protons is 176 keV, and for 63 MeV protons is 105 keV, after initially losing energy through the metal windows and masks. Since we observe no extra dark current for this array, we conclude that either the absorbed energy did not lead to substantial luminescence, or the energy was dissipated by other non-radiative processes through the crystal lattice. 

\begin{figure}[H]
\begin{center}
\begin{tabular}{c}
\includegraphics[height=7cm]{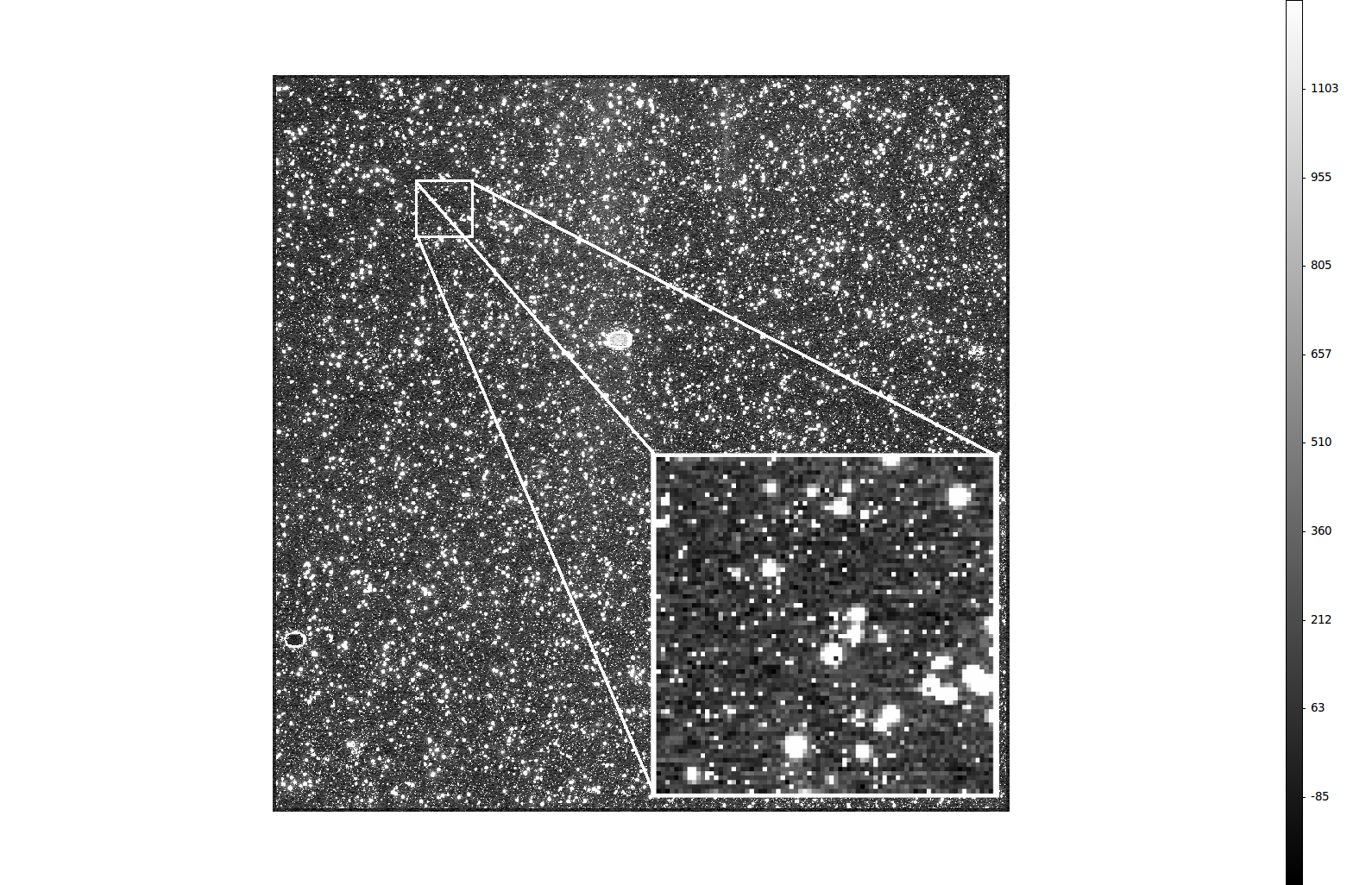}
\hspace{1cm}
\end{tabular}
\end{center}
\caption 
{ \label{fig:cds354}
Correlated double sample image of H1RG-17354 (an array with a 30 $\mu$m substrate) during 12 MeV irradiation integrated for 327 seconds. The scale on the right side of the image is in electrons. The large multi-pixel blotches (clusters) result from individual proton hits.  The small fraction of isolated hot pixels were identified as inoperable before radiation testing.  In between these are the `non-hit' pixels showing dark current $\sim$ 118e$^-$/327s = 0.36 e$^-$/s.} 
\end{figure} 

\noindent
In order to further illustrate the signal in the non-hit pixels, a correlated double sample image obtained with a long integration time is shown in Fig.~\ref{fig:cds354}. A correlated double sample image is defined as an image obtained immediately after reset subtracted from an image obtained after integrating over some time period. The integration time is therefore the time difference between those images.

 After a total dose of 7.5 krad(Si) (1.5x the lifetime dose for NEOCam) was applied on 13 Aug 2014, the test dewar was returned to Ames, and kept at 35 K while we continued to take data. Although the test dewar was still `hot' (secondaries are detected from the dewar material), we searched for high dark current pixels that were consistent throughout the data frames. Excluding pixels that were inoperable before irradiation and transient high signals that could be mistaken for permanently damaged pixels, we found that for an applied bias of 250 mV, 1.05\% newly identified pixels had high dark current and at 150 mV bias, 0.51\% newly identified pixels had high dark current. This high dark current is consistent with bias dependent tunneling current due to dislocations induced by bombardment of the protons. The dark current operability of the remainder of pixels was still \textgreater 90\%, and the modal dark current is unchanged within the measurement uncertainty. Because the modal dark current did not significantly change after the array was irradiated, we attribute the increase in baseline dark current from 0.04 e$^-$/s/pixel obtained on 13 Aug 2014 to 0.32 e$^-$/s/pixel obtained on 2 Oct 2014 (Table~\ref{table:darkcurrent354}) to a small light leak in the Ames dewar. These baseline dark currents are well below the NEOCam dark current specification (Table~\ref{table:neocamrequirements}). Although the plots in Fig~\ref{fig:darkcurrent354} are labeled as dark current, the current measured is the photocurrent from the light leak or glow plus the dark current. Hereafter, we refer to this level as `dark' current.

\subsubsection{H1RG-17346}

Data obtained with H1RG-17346 (substrate-removed) were reduced using similar methods as those employed for  H1RG-17354. We present results for the best region on the array (one quarter of the array). The FWHM for a histogram of the dark current data is greater than for the data presented for H1RG-17354 because these data are SUTR-16, rather than SUTR-64, and therefore have a factor of four fewer SUTR samples. Additionally, we only have a baseline measurement before the array was irradiated for an applied bias of 250 mV. Therefore we compare `dark' current before and during irradiation of SUTR-16 data with an applied bias of 250 mV.

\begin{figure}[H]
\begin{center}
\begin{tabular}{c}
\includegraphics[height=5.5cm]{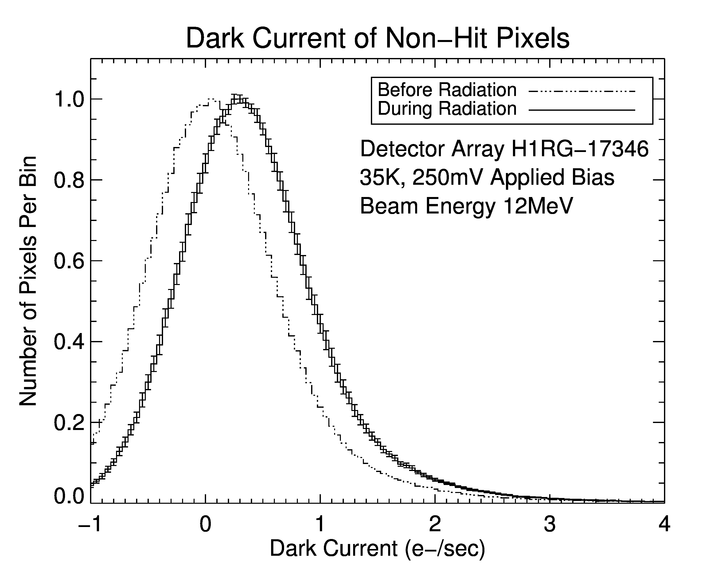}
\hfill
\includegraphics[height=5.5cm]{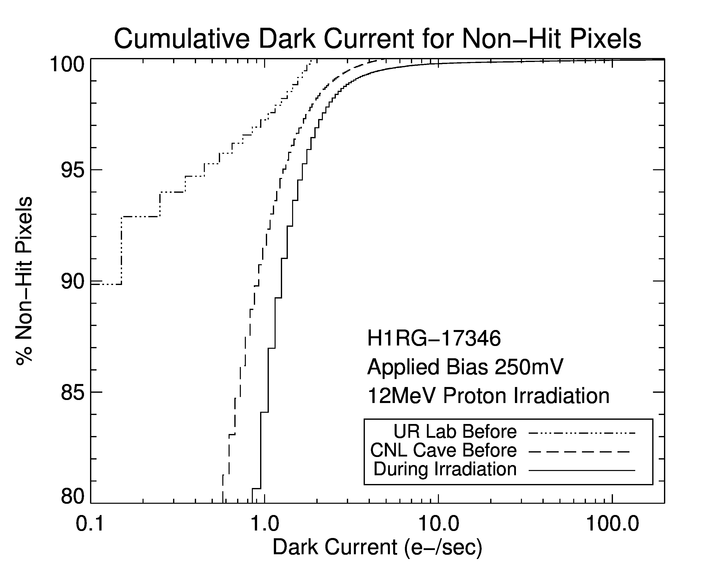}
\hspace{1cm}
\end{tabular}
\end{center}
\caption 
{ \label{fig:darkcurrent346}
 Left: `Dark' current for non-hit pixels during low dose 12 MeV proton irradiation for $\sfrac{1}{4}$ the area of H1RG-17346, the substrate-removed array. The `dark' current obtained at CNL with the beam stop in place before radiation testing is overplotted on the radiation data. Right: Cumulative `dark' current of non-hit pixels during 12 MeV proton irradiation. Cumulative dark current of the same pixels, tested in the lab at University of Rochester following exposure to a cumulative lifetime dose of radiation is overplotted, as is the cumulative `dark' current in the laboratory space outside the beam chamber at CNL immediately before irradiation. The applied reverse bias for these data is 250mV.} 
\end{figure} 

\begin{table}[H]
\centering
\caption{Summary of the modal `dark' current outside of the beam chamber before irradiation and during irradiation for various beam energies for H1RG-17346, the substrate-removed array.}
\begin{tabular}{lc} \toprule
    \multicolumn{1}{c}{\textbf{Beam Energy (MeV)}} & \multicolumn{1}{c}{\textbf{Modal Dark Current (e$^-$/s/pixel) }}  \\
    \multicolumn{1}{c}{\textbf{(MeV)}} & \multicolumn{1}{c}{\textbf{250 mV Reverse Bias}} \\  \midrule
    12 \textit{(and before irradiation)} & 0.33 $\pm$ 0.48 (\textit{0.14}) \\
    32 \textit{(and before irradiation)} & 0.14 $\pm$ 0.48 (\textit{0.14}) \\
    63 \textit{(and before irradiation)} & 0.38 $\pm$ 0.50 (\textit{0.14})  \\ \bottomrule
\end{tabular} 
\label{table:darkcurrent346}   
\end{table}

\noindent
The uncertainty in the measurement of `dark' current during this radiation experiment was higher than for the measurements of the previous detector array, and the perceived elevation in the modal current in non-hit pixels in the fully substrate-removed array (Table~\ref{table:darkcurrent346}) is within the uncertainties in the measurements.

\begin{figure}[H]
\begin{center}
\begin{tabular}{c}
\includegraphics[height=7cm]{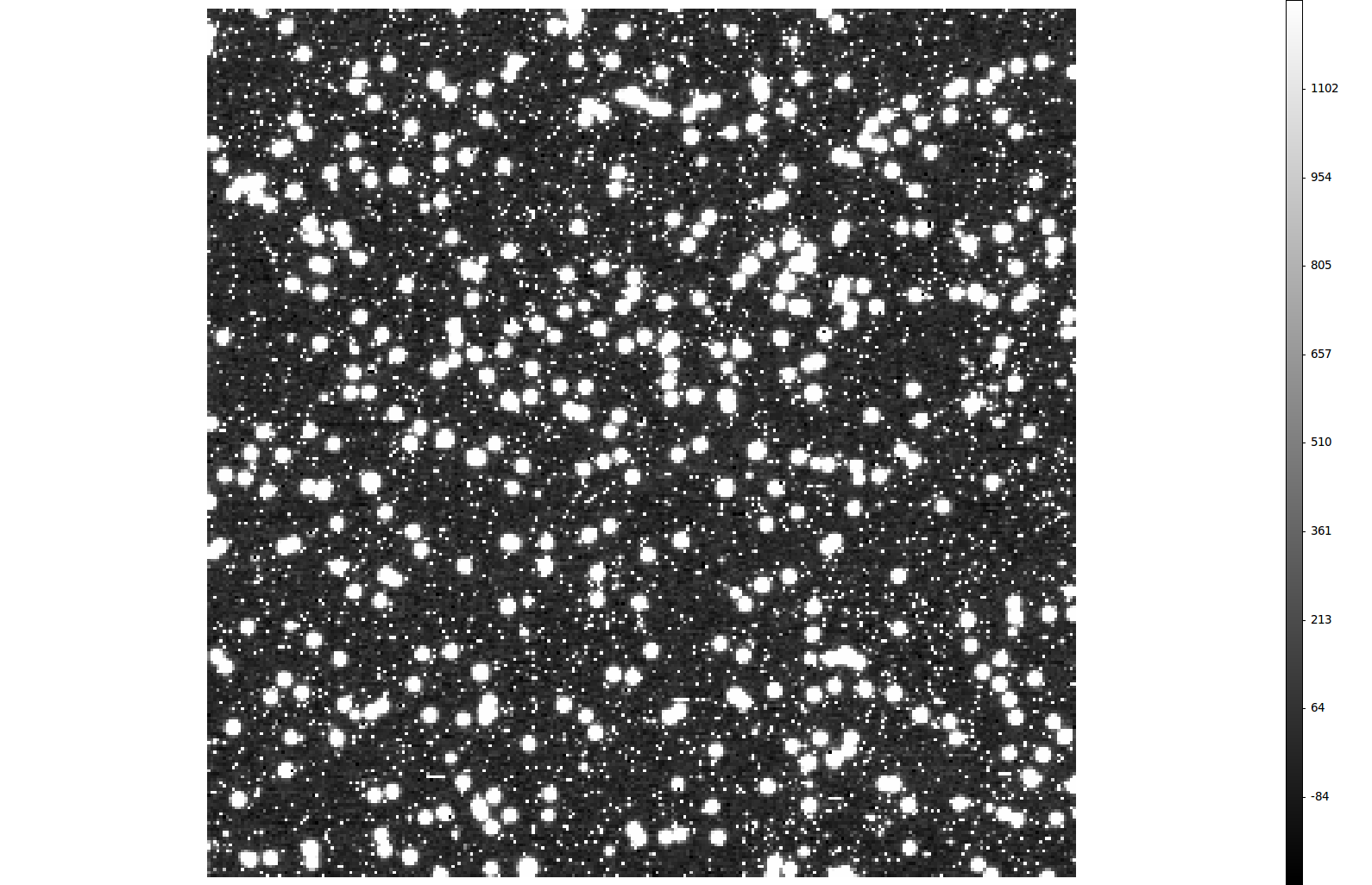}
\hspace{1cm}
\end{tabular}
\end{center}
\caption 
{ \label{fig:cds346}
Correlated double sample image of H1RG-17346, the substrate-removed array. This image was obtained by integrating for 74 sec while the array was being irradiated with 12 MeV protons. The applied bias for this integration was 250 mV.} 
\end{figure} 

\noindent
Following a cumulative lifetime dose of radiation at CNL, H1RG-17346 was warmed up and sent back to UR, where all performance characteristics were retested. For pixels that were operable before particle irradiation there is no change in modal dark current within the measurement uncertainty after the cumulative lifetime dose, although the operability decreased by a fraction of $\sim$1 percent for 150 mV applied bias and a few percent for 250 mV applied bias. 

\subsubsection{H1RG-16886}
\noindent
The final array tested, H1RG-16886 (full substrate), was irradiated with 18.1 and 34.4 MeV protons that differ from the energies used to test the two previously discussed arrays. The first energy, 18.1 MeV, was chosen to be as close as we were able to tune to 15.7 MeV, the energy at which the Bragg peak resides just within the 800 $\mu$m substrate. A beam energy of 34.4 MeV was chosen to be close to the 32 MeV beam energy utilized for the previous arrays tested.

\begin{figure}[H]
\begin{center}
\begin{tabular}{c}
\includegraphics[height=7cm]{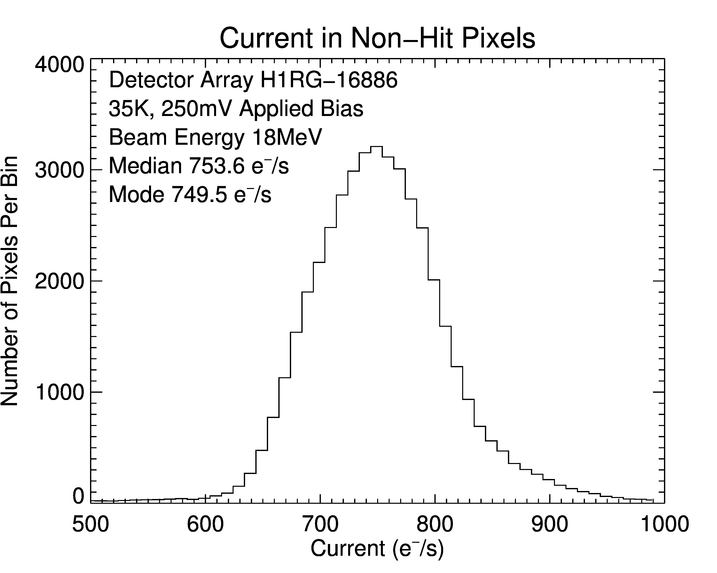}
\hspace{1cm}
\end{tabular}
\end{center}
\caption 
{ \label{fig:darkcurrent886}
Photocurrent for non-hit pixels during low dose 18.1 MeV proton irradiation for H1RG-16886. Data were obtained with 250 mV applied bias. } 
\end{figure} 

\begin{figure}[H]
\begin{center}
\begin{tabular}{c}
\includegraphics[height=7cm]{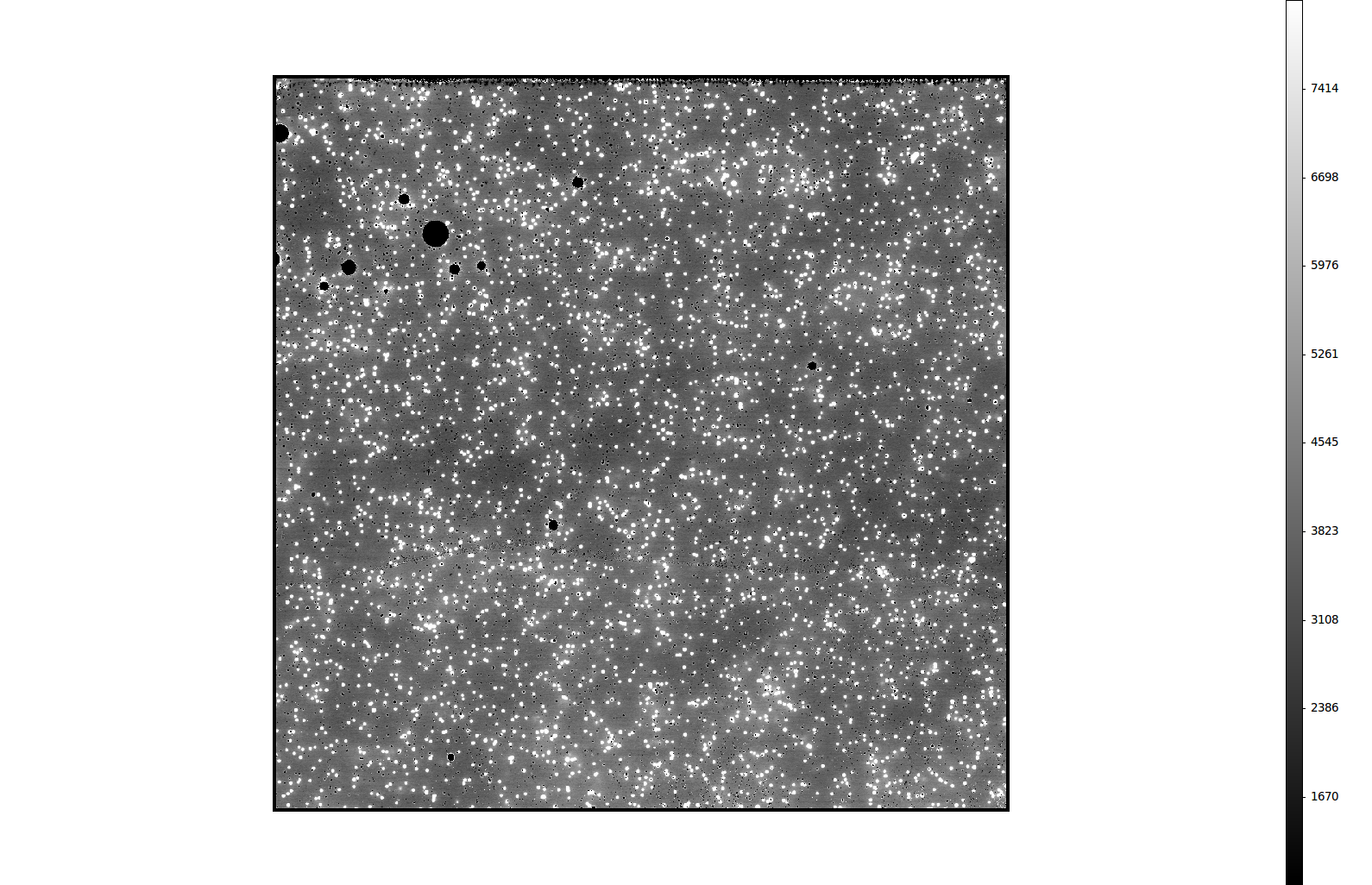}
\hspace{1cm}
\end{tabular}
\end{center}
\caption 
{ \label{fig:cds886}
Correlated double sample image of H1RG-16886, the array with 800 $\mu$m substrate, during 18.1 MeV irradiation integrated for 5.3 seconds with an applied bias of 250 mV. The scale on the right side of the image is in electrons. The flux level for these data is the same as in Table~\ref{table:darkcurrent886}.} 
\end{figure}

\noindent
The elevated current in non-hit pixels that we measure is 2-3 orders of magnitude above the dark current plus assumed light leak measured before irradiation. To further illustrate the luminescence in non-hit pixels, a 5.3 second integration time correlated double sampled image obtained with a beam energy of 18.1 MeV, and 250 mV applied bias is shown in Fig.~\ref{fig:cds886}.

\begin{table}[H]
\centering
\caption{Modal `dark' current in the laboratory space outside the beam chamber before irradiation and the background photocurrent detected during irradiation for various beam energies for H1RG-16886. The proton flux during irradiation by 18.1 MeV protons is 103 $\pm$ 20 protons/cm$^2$-s.}
\begin{tabular}{lll} \toprule
    \multicolumn{1}{c}{\textbf{Beam Energy }} & \multicolumn{2}{c}{\textbf{Modal Dark Current (e$^-$/s/pixel) }}  \\ \cmidrule(r){2-3}
    \multicolumn{1}{c}{\textbf{(MeV)}}              & \multicolumn{1}{c}{\textbf{150 mV Reverse Bias}} & \multicolumn{1}{c}{\textbf{250 mV Reverse Bias}} \\  \midrule
    18.1 \textit{(and before irradiation)} & 706 $\pm$ 22 \textit{(0.3)} & 749 $\pm$ 10 \textit{(0.4)} \\
    34.4 \textit{(and before irradiation)} & 81 $\pm$ 2 \textit{(0.3)} & 66 $\pm$ 2 \textit{(0.4)} \\\bottomrule
\end{tabular} 
\label{table:darkcurrent886}   
\end{table}

\noindent
The results in Table~\ref{table:darkcurrent886} are for a proton beam flux of 103 protons/cm$^2$-s. For the 5 $\mu$m cutoff wavelength InSb arrays on the Spitzer Space Telescope, which had pixels that were 30 $\mu$m in size and an array size of 256 x 256 (Area=0.59 cm$^2$), the rate of pixels hit was 3-10 s$^{-1}$ (Ref.~\citenum{hora2006}; Hora - private communication). Typically, they saw a single cosmic ray (CR) affecting between 2-4 pixels, smaller than what we report in Sec.~\ref{sect:hits}, because the pixel size for Spitzer is larger. This translates to 1-5 CR/s for the Spitzer arrays. To directly compare with the 103 protons/cm$^2$-s from the CNL beam, then the total integrated cosmic ray rate from Spitzer is 1.7 - 8.5 CR/cm$^2$-s, a factor of 12 lower. This reduces the $\sim$ 750 e$^-$/s luminescence we measure in the lab at CNL to \textbf{at most} 62 e$^-$/s in an ambient space-like environment, except when irradiated by solar flares. Since this luminescence current would be spatially and temporally variable, the 800 $\mu$m CdZnTe substrate might adversely affect the performance of NEOcam's detector arrays.


Testing of H1RG-16886 in the UR lab after radiation exposure and after an anneal to room temperature showed that the operability decreased by 1.2\% for an applied bias of 150 mV and by 2.7\% for an applied bias of 250 mV. 

\subsection{Single Proton Hits}
\label{sect:hits}
We examined the clusters of pixels corresponding to a single proton hit, and found distinct differences in both the charge distribution and the number of pixels associated with a hit as a function of substrate width, and beam energy. We first examine single proton hit data for the array with 30 $\mu$m CdZnTe substrate, H1RG-17354.

\begin{figure}[H]
\begin{center}
\begin{tabular}{c}
\includegraphics[height=5.5cm]{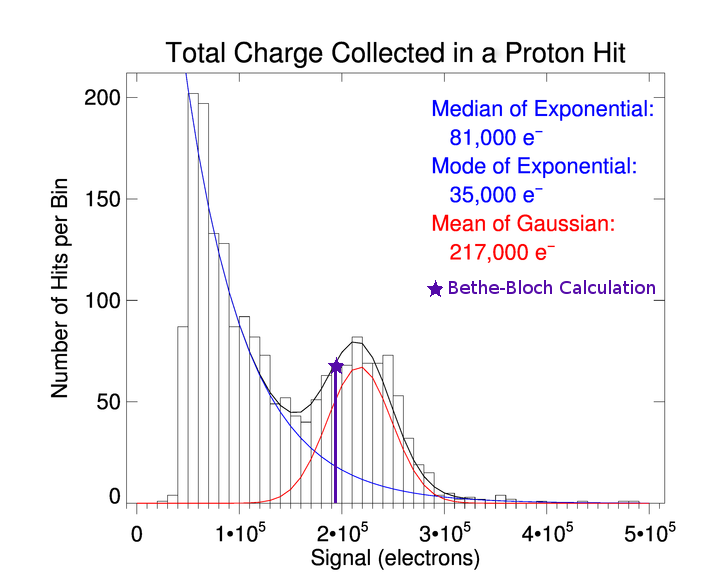}
\hfill
\includegraphics[height=5.5cm]{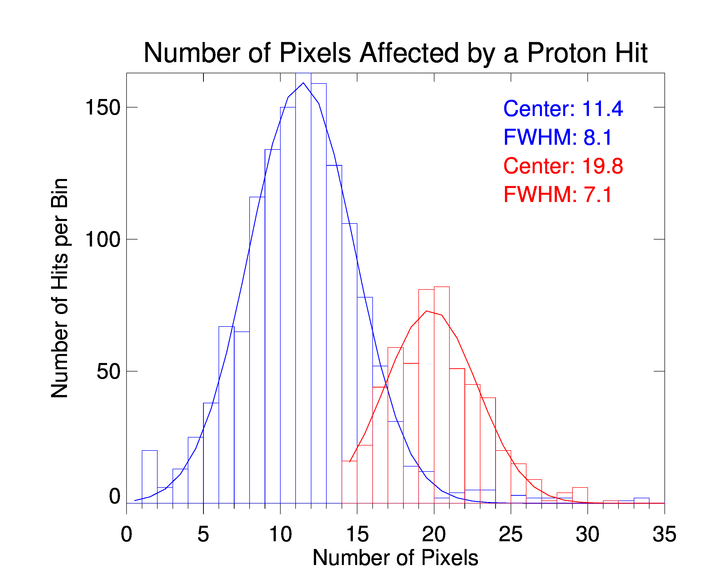}
\hspace{1cm}
\end{tabular}
\end{center}
\caption 
{ \label{fig:chargeandnpix354_12MeV}
Left: Histogram of total charge collected per proton hit in H1RG-17354, an array with a 30 $\mu$m substrate (Binsize = 10$^4$ e$^-$). Right: Histogram representing number of pixels affected by a single proton hit. The data shown here were obtained with an applied bias of 250 mV while the array is irradiated with 12 MeV protons.} 
\end{figure} 

\noindent
These data are background subtracted (SUTR-64) consecutive differences. The total hit is a sum of all pixels in the difference frame. The distribution in charge was bimodal, with separation in bimodality larger at 12 MeV energy than at the higher beam energies. We used an exponential plus Gaussian fit to model the charge distribution data: the exponential is not anticipated to be physical, but likely represents a portion of a second Gaussian (see Fig.~\ref{fig:chargeandnpix354_12MeV}, left, for a representative plot). These two distributions are fit to the data above 50,000 e$^-$ by a least squares method using the following model,

\begin{equation}
y_{fit} = Ae^{cx}+Be^{-(x-x_{center})/w^2}
\end{equation}

\noindent
where the parameters $A$ and $c$ describe the shape of an exponential distribution, and the parameters $B$, $x_{center}$, and $w$ describe the amplitude, center, and width of a Gaussian distribution. A possible explanation as to the origin of the distributions follows.

On the right in Fig~\ref{fig:chargeandnpix354_12MeV}, the total number of pixels affected by a proton hit is shown for the data on the left. SRIM simulations show the beam traveling directly through both the substrate and the detector layer. The spread in beam energies is not expected to produce two distinct distributions, so it is possible that the smaller number of pixels per hit, arising from the `exponential distribution' of charge collected per proton hit, were in fact not caused by protons, but by secondary particles and radiation. When the protons pass through the material surrounding the detector array, a nuclear reaction can occur that releases secondary particles and radiation. An atom can become activated, decay, and release secondaries. These secondaries are also reduced in flux when less beam attenuation is used to produce the higher beam energies. The purple line drawn on the charge distribution in Fig.~\ref{fig:chargeandnpix354_12MeV} indicates the charge calculated via the Bethe-Bloch formula assuming the 12 MeV proton beam energy was attenuated to $\sim$ 10.7 MeV before entering the back surface of the detector. After passage through the 30 $\mu$m CdZnTe, in this example, the beam loses $\sim$ 394 keV and travels 12 $\mu$m through the HgCdTe to the front surface of the detector, where the junction and depletion region are located.\cite{Nakamura2010} 


\begin{table}[H]
\centering
\caption{Summary of mean charge collected per proton hit and number of pixels per proton hit cluster at various beam energies for H1RG-17354, an array with a 30 $\mu$m substrate.  }
\begin{tabularx}{\linewidth}{lXXXX} \toprule
     Beam Energy & Mean total charge per proton hit (e$^-$) & Mean Number of pixels per cluster & Mean total charge per proton hit (e$^-$) & Mean Number of pixels per cluster  \\ \cmidrule(r){2-3} \cmidrule(r){4-5}
    \multicolumn{1}{c}{\textbf{(MeV)}}  & \multicolumn{2}{c}{\textbf{150 mV Reverse Bias}} & \multicolumn{2}{c}{\textbf{250 mV Reverse Bias}} \tabularnewline \midrule
    12 & 295,000 & 24 $\pm$ 4 & 217,000 & 20 $\pm$ 4  \\ 
    32 & 98,000   & 14 $\pm$ 2 & 67,000   & 12 $\pm$ 2  \\
    63 & 86,000   & 19 $\pm$ 3 & 64,000   & 12 $\pm$ 3 \\ \bottomrule
\end{tabularx} 
\label{table:summarycharge354}   
\end{table}

\begin{table}[H]
\centering
\caption{Bethe-Bloch calculation for energy loss in 12 $\mu$m thick HgCdTe after passage through 30 $\mu$m CdZnTe}
\begin{tabularx}{\linewidth}{cccc} \toprule
  \multicolumn{1}{c}{\textbf{Beam Energy }} & 
  \multicolumn{1}{c}{\textbf{HgCdTe $dE/dx$ }}  &
  \multicolumn{1}{c}{\textbf{Total Calculated}} & 
  \multicolumn{1}{c}{\textbf{Mean of Total }} \tabularnewline
      \multicolumn{1}{c}{\textbf{(MeV)}} & 
      \multicolumn{1}{c}{\textbf{(keV/$\mu$m)}}  &
      \multicolumn{1}{c}{\textbf{Charge (e$^-$)}} &
       \multicolumn{1}{c}{\textbf{Observed Proton Hits (e$^-$)}}   \tabularnewline 
      \multicolumn{1}{c}{\textbf{}} & 
      \multicolumn{1}{c}{\textbf{}}  &
      \multicolumn{1}{c}{\textbf{}} &
       \multicolumn{1}{c}{\textbf{150 mV bias, 250 mV bias}}   \tabularnewline  \midrule     
	12	&	16.7	 &	193,000  & 217,000, 295,000\\
	32	&      7.59	 &	88,000    & 67,000, 98,000 \\
	63	&	4.55  & 	52,000    & 64,000, 86,000 \\ \bottomrule
\end{tabularx} 
\label{table:bethe354}   
\end{table}

The mean of the Gaussian charge distribution for all three beam energies is listed in Table~\ref{table:summarycharge354} for H1RG-17354, corresponding to different numbers of pixels affected by a proton hit.

Table~\ref{table:bethe354} shows the estimated charged generated in the HgCdTe detector layer after first losing energy through the metal windows and masks in front of the dewar and then through the 30 $\mu$m CdZnTe from Eqs.~\ref{eq:energytransfer} and~\ref{eq:chargegen}. These values can be compared to the mean total charge for the observed proton hits (column 4). The two values listed for the proton hits we observe correspond to the two applied biases utilized for testing. Similar results for H1RG-17346 (substrate-removed) and H1RG-16886 (full substrate) are shown in Tables~\ref{table:summarycharge346} through~\ref{table:bethe886} below.

\begin{table}[H]
\centering
\caption{Summary of mean charge collected per proton hit and number of pixels per proton hit cluster at various beam energies for H1RG-17346, the substrate-removed array. Larger charge events affect more pixels. These cluster sizes as a function of proton energy are similar to those seen for device H1RG-17354 (the array with a 30 $\mu$m substrate).}
\begin{tabularx}{\linewidth}{lXXXX} \toprule
     Beam Energy & Mean total charge per proton hit (e$^-$) & Mean Number of pixels per cluster & Mean total charge per proton hit (e$^-$) & Mean Number of pixels per cluster  \\ \cmidrule(r){2-3} \cmidrule(r){4-5}
    \multicolumn{1}{c}{\textbf{(MeV)}}  & \multicolumn{2}{c}{\textbf{150 mV Reverse Bias}} & \multicolumn{2}{c}{\textbf{250 mV Reverse Bias}} \tabularnewline \midrule
    12 & 238,000 & 25 $\pm$ 5 & 238,000 & 21 $\pm$ 5 \\ 
    32 & 85,000   & 17 $\pm$ 3 & 84,000   & 14 $\pm$ 3  \\
    63 & 65,000   & 12 $\pm$ 4 & 65,000   & ~~9 $\pm$ 4 \\ \bottomrule
\end{tabularx} 
\label{table:summarycharge346}   
\end{table}

\begin{table}[H]
\centering
\caption{Charge deposition for H1RG-17346, the substrate-removed array.}
\begin{tabularx}{\linewidth}{cccc} \toprule
  \multicolumn{1}{c}{\textbf{Beam Energy }} & 
  \multicolumn{1}{c}{\textbf{HgCdTe $dE/dx$ }}  &
  \multicolumn{1}{c}{\textbf{Total Calculated}} & 
  \multicolumn{1}{c}{\textbf{Mean of Total }} \tabularnewline
      \multicolumn{1}{c}{\textbf{(MeV)}} & 
      \multicolumn{1}{c}{\textbf{(keV/$\mu$m)}}  &
      \multicolumn{1}{c}{\textbf{Charge (e$^-$)}} &
       \multicolumn{1}{c}{\textbf{Observed Proton Hits (e$^-$)}}   \tabularnewline
        \multicolumn{1}{c}{\textbf{}} & 
      \multicolumn{1}{c}{\textbf{}}  &
      \multicolumn{1}{c}{\textbf{}} &
       \multicolumn{1}{c}{\textbf{150 mV bias, 250 mV bias}}   \tabularnewline  \midrule       
    12 & 16.7 & 193,000 & 238,000, 238,000 \\
    32 & 7.60 & 88,000   & 84,000, 85,000  \\
    63 & 4.55 & 53,000   & 65,000, 65,000 \\ \bottomrule
\end{tabularx} 
\label{table:bethe346}   
\end{table}

\begin{table}[H]
\centering
\caption{Summary of median charge collected per proton hit and number of pixels per proton hit cluster at various beam energies for H1RG-16886, the array with full 800 $\mu$m substrate. The smaller proton hit size corresponding to a larger total charge per proton hit is different from the previous observations with the 30 $\mu$m substrate and the array with no substrate.}
\begin{tabularx}{\linewidth}{lXXXX} \toprule
     Beam Energy & Mean total charge per proton hit (e$^-$) & Mean Number of pixels per cluster & Mean total charge per proton hit (e$^-$) & Mean Number of pixels per cluster  \\ \cmidrule(r){2-3} \cmidrule(r){4-5}
    \multicolumn{1}{c}{\textbf{(MeV)}}  & \multicolumn{2}{c}{\textbf{150 mV Reverse Bias}} & \multicolumn{2}{c}{\textbf{250 mV Reverse Bias}} \tabularnewline \midrule
    18.1 & 184,000 & ~~7 $\pm$ 3 & 245,000 & ~~7 $\pm$ 5 \\ 
    34.4 & 104,000 & 11 $\pm$ 4 & 112,000 & 11 $\pm$ 4  \\  \bottomrule
\end{tabularx} 
\label{table:summarycharge886}   
\end{table}

\begin{table}[H]
\centering
\caption{Bethe-Bloch calculation for H1RG-16886 - 800 $\mu$m CdZnTe and 12 $\mu$m thick HgCdTe detector configuration.}
\begin{tabularx}{\linewidth}{cccc} \toprule
  \multicolumn{1}{c}{\textbf{Beam Energy }} & 
  \multicolumn{1}{c}{\textbf{HgCdTe $dE/dx$ }}  &
  \multicolumn{1}{c}{\textbf{Total Calculated}} & 
  \multicolumn{1}{c}{\textbf{Mean of Total }} \tabularnewline
      \multicolumn{1}{c}{\textbf{(MeV)}} & 
      \multicolumn{1}{c}{\textbf{(keV/$\mu$m)}}  &
      \multicolumn{1}{c}{\textbf{Charge (e$^-$)}} &
       \multicolumn{1}{c}{\textbf{Observed Proton Hits (e$^-$)}}   \tabularnewline 
             \multicolumn{1}{c}{\textbf{}} & 
      \multicolumn{1}{c}{\textbf{}}  &
      \multicolumn{1}{c}{\textbf{}} &
       \multicolumn{1}{c}{\textbf{150 mV bias, 250 mV bias}}   \tabularnewline  \midrule        
    18.1 & 11.9 & 138,000    & 184,000, 245,000 \\
    34.4 & 7.19   & 83,000   & 104,000, 112,000 \\ \bottomrule
\end{tabularx} 
\label{table:bethe886}   
\end{table}

\noindent
The proton beam loses energy as it first passes through the window and metal masks in the beam path, then further as it ionizes material along its path, in both the CdZnTe substrate and the HgCdTe material. For charge spreading across multiple pixels from a single proton hit, the dominant mechanisms are coulomb repulsion and enhanced diffusion within the bulk material. Inter pixel capacitance (IPC) is only responsible for $\sim$3 pixels of charge spread.\cite{McMurtry2004} In the HgCdTe material, holes associated with a given proton hit  from the electron-hole pairs produced at the incident surface will diffuse more than those produced close to a junction, and overall will result in a spread in the charge detected by the diode. This is similar to the loss of image quality or lower MTF seen as the wavelength of light is changed from longer to shorter, i.e. the long wavelength photons are typically absorbed near the p-n junction after they pass through most of the bulk material, while the short wavelength photons are absorbed near the backside surface or furthest away from the p-n junction, hence allowing for the greatest amount of diffusion spread.

Tables~\ref{table:summarycharge354}, ~\ref{table:summarycharge346}, and~\ref{table:summarycharge886} show that for the lowest energy protons there is a variation in total hit size that is inversely related to the amount of substrate on a given device. The enhanced charge spreading seen in the low energy proton hits can be attributed to the loss of the charge (hole) due either to radiative or non-radiative, e.g. phonon, processes if that hole is allowed to enter the substrate.  A hole (charge carrier) may only enter the substrate if it has more than the bandgap energy of CdZnTe, which is about 1.6 eV.\cite{quijada2007} The hole mobility in CdZnTe is much lower (\textless 100 cm$^2$/V/s) than HgCdTe.  Thus the hole may be more easily lost (775 nm photon or thermalized via phonon) once it enters the CdZnTe.   In other words, although all of the charge deposited by a proton will be initially located in a narrow column along that proton track, the charges (holes) will spread laterally due to coulomb repulsion and diffusion.  All of those charges (holes) will be able to diffuse into 4$\pi$ steradians unless there is a boundary such as the p-n junction or the HgCdTe to vacuum boundary, in which case the charge (hole) may only diffuse into 2$\pi$ steradians, i.e. reflect back in the direction of a p-n junction.  Hence for the case of an array without a substrate, the holes closest to the HgCdTe to vacuum boundary will be reflected from that surface and produce, on average, a larger amount of lateral diffusion.  For an array with a substrate, the holes may be lost if they enter the substrate. 

However, for the higher energy proton irradiation, we find almost no change in pixel hit size from one device to the next.  This variation in charge spreading versus energy of the incident protons is again related to the total amount of charge initially deposited and thus related to the amount of coulomb repulsion that would occur for the given amount of charge for the same volume.

\begin{figure}[H]
\begin{center}
\begin{tabular}{c}
\includegraphics[height=4.8cm]{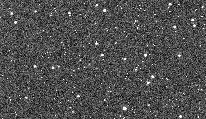}
\hfill
\includegraphics[height=4.8cm]{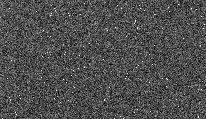}
\hspace{1cm}
\end{tabular}
\end{center}
\caption 
{ \label{fig:beforeafterreset354}
Subarray image of a region on the array at the end of a 169 second integration (left), and the same region of the array immediately after reset (right).} 
\end{figure}

After the array has been hit with protons while we integrate charge, the array immediately recovers after reset (Fig.~\ref{fig:beforeafterreset354}). Note that the recovered array image shows a typical dark response with the occasional hot and dark pixels. Our results confirm those of Girard et al. (2014), who showed that a pixel's dark current, responsivity, and noise were unaffected 5.5 seconds after a cosmic ray hit.

\noindent
No significant latent images from the proton hits were observed in the frames obtained immediately after frames exposed to individual hits.

\section{Summary}

Proton testing on NEOCam developed 10 $\mu$m cutoff arrays was conducted in order to assess whether we could detect luminescence, examine any residual effects and recovery time from proton hits, and assess the array performance following a cumulative lifetime dose of irradiation. 

The `dark' current operability of non-hit pixels during each frame of low fluence proton irradiation of H1RG-17354 and H1RG-17346 was $\sim$ 100\%, fully meeting the NEOCam requirement of \textless 200 e$^-$/sec. Proton induced luminescence is not significant after the removal of $\sim$ 95\% of the CdZnTe substrate for our LWIR arrays, H1RG-17354. The observed `dark' current is unchanged within measurement uncertainties. On the other hand, H1RG-16886, the array with a fully intact CdZnTe substrate did lead to a substantial luminescence for a flux level of 103 protons/cm$^2$-s at a beam energy of 18.1 MeV. 

Both H1RG-17354 and H1RG-17346 were subjected to a cumulative lifetime dose of at least 5 krad(Si). Lab testing at UR post-total lifetime dose irradiation conducted on H1RG-17346 (after the array was warmed up and cooled down again) showed that the dark current and well depth operability decreased by a fraction of a percent with an applied bias of 150 mV, and with an applied bias of 250 mV, the operability had decreased a few percent (Table~\ref{table:neocamrequirements}). For both applied biases, the detector array still meets NEOCam operability requirements after a lifetime dose of radiation.

Measurements of the dark current of one array, H1RG-17354, immediately after a lifetime dose of proton irradiation showed that at most $\sim$ 1\% of pixels sustained permanent damage that led to substantial dark current. For our LWIR 10 $\mu$m cutoff HgCdTe detector arrays, we find that they will still meet the NEOCam dark current and operability requirements after a lifetime dose of protons.
 
For pixels hit by individual protons, the short-term effect is transient. The charge is spread out over $\sim$ 10 pixels for higher energy protons, and up to 21 pixels for lower energy protons hitting H1RG-17346 and H1RG-17354, and $\sim$ 10 pixels for low-energy protons hitting H1RG-16886. A typical hit is well below saturation, so data before and after the event are usable.  

\appendix   
\section{Dark Current Histograms}
\label{sect:AppendixA}

\begin{figure}[H]
\begin{center}
\begin{tabular}{c}
\includegraphics[height=9cm]{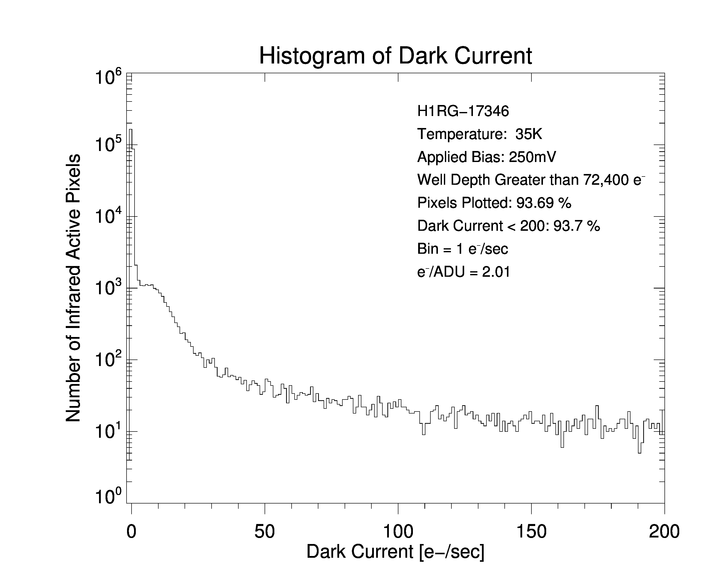}
\hspace{1cm}
\end{tabular}
\end{center}
\caption 
{ \label{fig:operabilitybefore346}
Histogram of dark current for pixels with well depth \textgreater 72,400 e$^-$ for H1RG-17346, the substrate-removed array.} 
\end{figure} 

\begin{figure}[H]
\begin{center}
\begin{tabular}{c}
\includegraphics[height=9cm]{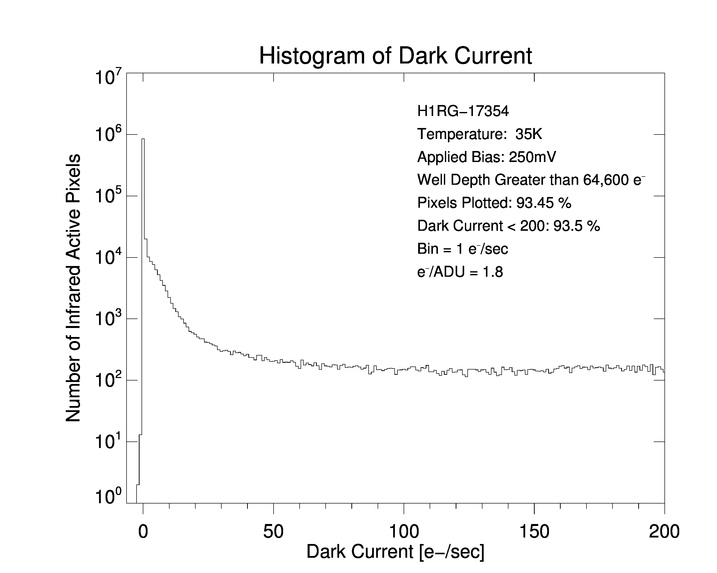}
\hspace{1cm}
\end{tabular}
\end{center}
\caption 
{ \label{fig:operabilitybefore354}
Histogram of dark current for pixels with well depth \textgreater 64,600 e$^-$ for H1RG-17354, the array with 30 $\mu$m substrate.} 
\end{figure} 

\begin{figure}[H]
\begin{center}
\begin{tabular}{c}
\includegraphics[height=9cm]{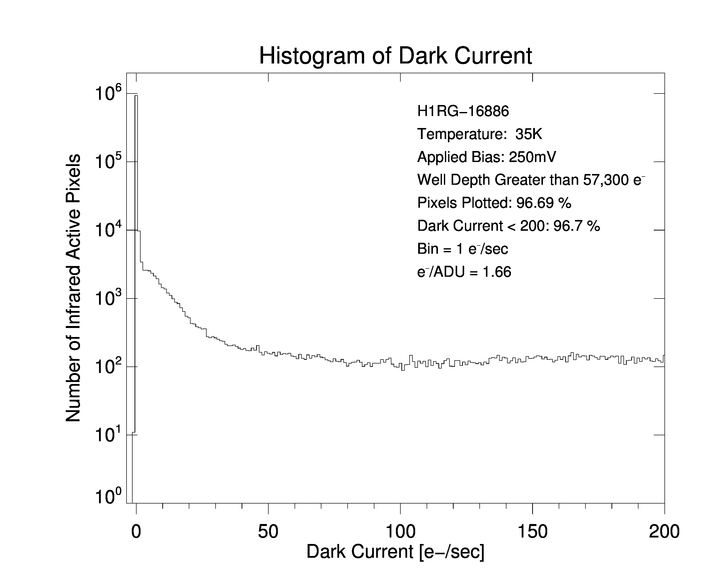}
\hspace{1cm}
\end{tabular}
\end{center}
\caption 
{ \label{fig:operabilitybefore886}
Histogram of dark current for pixels with well depth \textgreater 57,300 e$^-$ for H1RG-16886, the array with 800 $\mu$m substrate.} 
\end{figure}

\acknowledgments 
The research was carried out at the University of Rochester, the NASA Ames Research Center, the University of California Davis Crocker Nuclear Laboratory, and the Jet Propulsion Laboratory, California Institute of Technology, under contracts with the National Aeronautics and Space Administration.  

Special thanks to Frank Wolfs for helpful feedback on this paper.

\bibliography{Dorn2016_Manuscript}   

\begin{thebibliography}{10}

\bibitem{Mainzer15}
A.~Mainzer, T.~Grav, J.~Bauer, T.~Conrow, R.~M. Cutri, J.~Dailey, J.~Fowler,
  J.~Giorgini, T.~Jarrett, J.~Masiero, T.~Spahr, T.~Statler, and E.~L. Wright,
  ``Survey simulations of a new near-earth asteroid detection system,'' {\em
  The Astronomical Journal} {\bf 149}(5), 172  (2015).

\bibitem{Rauscher2014}
B.~J. Rauscher, N.~Boehm, S.~Cagiano, G.~S. Delo, R.~Foltz, M.~A. Greenhouse,
  M.~Hickey, R.~J. Hill, E.~Kan, D.~Lindler, D.~Mott, A.~Waczynski, and Y.~Wen,
  ``New and better detectors for the {JWST} {Near-Infrared Spectrograph},''
  {\em Publications of the Astronomical Society of the Pacific} {\bf 126}(942),
  739  (2014).

\bibitem{McMurtry2013}
C.~W. McMurtry, D.~Lee, J.~Beletic, C.~A. Chen, R.~T. Demers, M.~Dorn,
  D.~Edwall, C.~B. Fazar, W.~J. Forrest, F.~Liu, A.~K. Mainzer, J.~L. Pipher,
  and A.~Yulius, ``Development of sensitive long-wave infrared detector arrays
  for passively cooled space missions,'' {\em Optical Engineering} {\bf 52}(9),
  091804--091804  (2013).

\bibitem{Oday2011}
J.~O. Day, C.~W. OÕDell, R.~Pollock, C.~J. Bruegge, D.~Rider, D.~Crisp, and
  C.~E. Miller, ``Preflight spectral calibration of the {Orbiting Carbon
  Observatory},'' {\em IEEE transactions on geoscience and remote sensing} {\bf
  49}(7), 2793--2801  (2011).

\bibitem{Mainzer2008}
A.~Mainzer, M.~Larsen, M.~G. Stapelbroek, H.~Hogue, J.~Garnett, M.~Zandian,
  R.~Mattson, S.~Masterjohn, J.~Livingston, N.~Lingner, N.~Alster, M.~Ressler,
  and F.~Masci, ``Characterization of flight detector arrays for the wide-field
  infrared survey explorer,'' in {\em High Energy, Optical, and Infrared
  Detectors for Astronomy III},  {\em Proc. SPIE} {\bf 7021},
  70210X--70210X--12  (2008).

\bibitem{Baggett2008}
S.~M. Baggett, R.~J. Hill, R.~A. Kimble, J.~W. MacKenty, A.~Waczynski, H.~A.
  Bushouse, N.~Boehm, H.~E. Bond, T.~M. Brown, N.~R. Collins, G.~Delo,
  L.~Dressel, R.~Foltz, G.~Hartig, B.~Hilbert, E.~Kan, J.~Kim-Quijano,
  E.~Malumuth, A.~Martel, P.~McCullough, L.~Petro, M.~Robberto, and Y.~Wen,
  ``The wide-field camera 3 detectors,'' in {\em High Energy, Optical, and
  Infrared Detectors for Astronomy III},  {\em Proc. SPIE} {\bf 7021},
  70211Q--70211Q--11  (2008).

\bibitem{Laureijs2012}
R.~Laureijs, P.~Gondoin, L.~Duvet, G.~S. Criado, J.~Hoar, J.~Amiaux, J.-L.
  Augu\`{e}res, R.~Cole, M.~Cropper, A.~Ealet, P.~Ferruit, I.~E. Sanz,
  K.~Jahnke, R.~Kohley, T.~Maciaszek, Y.~Mellier, T.~Oosterbroek, F.~Pasian,
  M.~Sauvage, R.~Scaramella, M.~Sirianni, and L.~Valenziano, ``Euclid: {ESA's}
  mission to map the geometry of the dark universe,'' in {\em Space Telescopes
  and Instrumentation 2012: Optical, Infrared, and Millimeter Wave},  {\em
  Proc. SPIE} {\bf 8442}, 84420T--84420T--8  (2012).

\bibitem{quijada2007}
M.~A. Quijada and R.~Henry, ``Temperature evolution of exciton absorptions in
  {Cd$_{1-x}$Zn$_x$Te} materials,'' in {\em Optical Engineering+ Applications},
   669206--669206, International Society for Optics and Photonics  (2007).

\bibitem{Piquette2008}
E.~C. Piquette, D.~D. Edwall, H.~Arnold, A.~Chen, and J.~Auyeung,
  ``Substrate-removed {HgCdTe}-based focal-plane arrays for short-wavelength
  infrared astronomy,'' {\em Journal of Electronic Materials} {\bf 37}(9),
  1396--1400  (2008).

\bibitem{Waczynski2005}
A.~Waczynski, P.~W. Marshall, C.~J. Marshall, R.~Foltz, R.~A. Kimble, S.~D.
  Johnson, and R.~J. Hill, ``Radiation induced luminescence of the {CdZnTe}
  substrate in {HgCdTe} detectors for {WFC3},'' in {\em Focal Plane Arrays for
  Space Telescopes II},  {\em Proc. SPIE} {\bf 5902}, 59020P--59020P--10
  (2005).

\bibitem{Smith2006}
R.~Smith, C.~Bebek, M.~Bonati, M.~G. Brown, D.~Cole, G.~Rahmer, M.~Schubnell,
  S.~Seshadri, and G.~Tarle, ``Noise and zero point drift in 1.7$\mu$m cutoff
  detectors for {SNAP},'' in {\em High Energy, Optical, and Infrared Detectors
  for Astronomy II},  {\em Proc. SPIE} {\bf 6276}, 62760R--62760R--16  (2006).

\bibitem{Garnett2004}
J.~D. Garnett, M.~C. Farris, S.~S. Wong, M.~Zandian, D.~Hall, S.~Jacobson,
  G.~Luppino, S.~Parker, D.~Dorn, S.~Franka, E.~Freymiller, and S.~McMuldroch,
  ``{2Kx2K} molecular beam epitaxy {HgCdTe} detectors for the {James Webb Space
  Telescope NIRCam instrument},'' in {\em Optical and Infrared Detectors for
  Astronomy},  {\em Proc. SPIE} {\bf 5499}, 35--46  (2004).

\bibitem{Girard2014}
J.~J. Girard, W.~J. Forrest, C.~W. McMurtry, J.~L. Pipher, M.~Dorn, and
  A.~Mainzer, ``Cosmic ray response of megapixel {LWIR} arrays from {TIS},'' in
  {\em High Energy, Optical, and Infrared Detectors for Astronomy VI},  {\em
  Proc. SPIE} {\bf 9154}, 91542A--91542A--10  (2014).

\bibitem{Nakamura2010}
K.~Nakamura and {\relax Particle Data Group}, ``Review of particle physics,''
  {\em Journal of Physics G: Nuclear and Particle Physics} {\bf 37}(7A), 075021
   (2010).

\bibitem{Bethe1930}
H.~Bethe, ``Zur theorie des durchgangs schneller korpuskularstrahlen durch
  materie,'' {\em Annalen der Physik} {\bf 397}(3), 325--400  (1930).

\bibitem{Marshall2003}
P.~W. Marshall, J.~E. Hubbs, D.~C. Arrington, C.~J. Marshall, G.~G. R.~A.~Reed,
  J.~C. Pickel, and R.~A. Ramos, ``Proton-induced transients and charge
  collection measurements in a {LWIR HgCdTe} focal plane array,'' {\em IEEE
  Transactions on Nuclear Science} {\bf 50}(6), 1968--1973  (2003).

\bibitem{Pickel2004}
J.~C. Pickel, R.~A. Reed, P.~W. Marshall, A.~Waczynski, E.~Polidan, S.~Johnson,
  R.~McMurray, M.~McKelvey, K.~Ennico, R.~Johnson, and G.~Gee,
  ``Radiation-induced transient effects in {HgCdTe IR} focal plane arrays,'' in
  {\em Optical, Infrared, and Millimeter Space Telescopes},  {\em Proc. SPIE}
  {\bf 5487}, 698--709  (2004).

\bibitem{Logachev2002}
Y.~I. Logachev, K.~Kecskem{\'e}ty, and M.~A. Zeldovich, ``Energy spectra of
  low-flux protons in the inner heliosphere under quiet solar conditions,''
  {\em Solar Physics} {\bf 208}(1), 141--166  (2002).

\bibitem{Simpson1983}
J.~A. Simpson, ``Elemental and isotopic composition of the galactic cosmic
  rays,'' {\em Annual Review of Nuclear and Particle Science} {\bf 33}(1),
  323--382  (1983).

\bibitem{srim}
``Software documentation and description for the stopping range of ions in
  matter.'' \url{http://www.srim.org}.

\bibitem{srimbook}
J.~F. Ziegler, J.~P. Biersack, and M.~D. Ziegler, {\em {SRIM} The Stopping and
  Range of Ions in Matter}, Lulu Press Co.  (2015).

\bibitem{hora2006}
J.~L. Hora, B.~M. Patten, G.~G. Fazio, and W.~J. Glaccum, ``The effects of
  cosmic rays and solar flares on the {IRAC} detectors: the first two years of
  in-flight operation,'' in {\em High Energy, Optical, and Infrared Detectors
  for Astronomy II},  {\em Proc. SPIE} {\bf 6276}, 62760J--62760J--15  (2006).

\bibitem{McMurtry2004}
C.~W. McMurtry, W.~J. Forrest, A.~C. Moore, and J.~L. Pipher, ``{James Webb
  Space Telescope: }characterization of flight candidate {NIR InSb} arrays,''
  {\em Proc. SPIE} {\bf 5167}, 144--158  (2004).

\end{thebibliography}
\bibliographystyle{spiejour}   


\vspace{2ex}\noindent\textbf{Meghan L. Dorn} is a Ph.D. student in materials science at the University of Rochester. She received her BS degree in imaging science from Rochester Institute of Technology and her MS degree in optics from the University of Rochester in 2012 and 2015, respectively. 

\vspace{1ex}
\noindent Biographies and photographs of the other authors are not available.


\listoffigures

\end{spacing}
\end{document}